# 3D-printed stand, timing interface, and coil localization tools for concurrent TMS-fMRI experiments

Samuel Goldstein, Farshad Rafiei, and Dobromir Rahnev

School of Psychology, Georgia Institute of Technology, Atlanta, GA

**Acknowledgements**

This work was supported by the National Institute of Health (award R21MH122825).

**Competing interests:**

None

**Correspondence**

Samuel Goldstein

Georgia Institute of Technology

654 Cherry Str. NW

Atlanta, GA 30332

E-mail: sgoldstein41@gatech.edu

**Abstract**

Concurrent TMS-fMRI involves administrating TMS while subjects are inside an MRI scanner and allows the study of the effects of neurostimulation on simultaneous brain activity. Despite its high promise, the technique has proven challenging to implement for at least three reasons. First, it is difficult to position and stabilize the TMS coil inside the MRI scanner in a way that precisely targets a pre-specified brain region. Second, standard task-presentation software suffers from imprecise timing, which can lead to TMS causing large image artifacts. Third, it is difficult to verify the exact TMS coil position during scanning. In this paper, we describe solutions to all three of these challenges. First, we develop a 3D-printed TMS stand that is fully adjustable and can reach most areas of the scalp. The stand is compatible with various MR coils and features an adjustable mirror holder. Second, we create an interface that can precisely time the TMS pulses with respect to the fMRI image acquisition with a variance of under 1 ms. Third, we develop software for precisely determining the TMS coil position inside the MRI scanner and computing the location of maximal stimulation. All three tools are either free or inexpensive. We provide detailed instructions for building and implementing these tools to facilitate an efficient and reliable concurrent TMS-fMRI setup.

## Introduction

Transcranial magnetic stimulation (TMS) is a neurostimulation tool that can influence neural excitability by generating a magnetic field which induces an electric field in the targeted region. TMS allows the investigation of causal relationships between localized neural activity and both behavioral and connectivity changes (Hallett, 2007), and can also be used for clinical therapies such as depression treatment (Loo & Mitchell, 2005; Paus & Barrett, 2004). However, the effects of TMS on brain activity are still not well understood, which has led researchers to combine TMS with imaging methods such as functional magnetic resonance imaging (fMRI) (Bergmann et al., 2021; Bestmann et al., 2008; Hawco et al., 2018; Ruff et al., 2006, 2008). The resulting technique – referred to as concurrent TMS-fMRI – was developed over 20 years ago (Bohning et al., 1997, 1998) and has the promise to uncover the neurophysiological effects of TMS and thus advance clinical and cognitive neuroscience research.

However, despite its promise, concurrent TMS-fMRI is yet to be widely adopted, which is likely caused by the fact that the technique is very challenging to implement. Below we discuss three specific challenges that have hampered research, review previous solutions to them, and introduce three new tools to address them. In later sections we provide more details around the setup and operation of each tool, and validate their utility.

Placing TMS coil inside MRI

One of the greatest challenges involved in concurrent TMS-fMRI is the appropriate placement of the TMS coil inside the MRI scanner. The difficulty is mostly due to the weight of the TMS coil and the need to work around the MRI receiver coils that a researcher may be using. Additional complications often arise when employing several MRI receiver coils to maximize the field of view and MR signal strength, as well as when attempting to place a mirror to allow the subject to view a screen.

Because the TMS coil cannot fit in standard MR coils for whole-brain scanning, previous studies have used various custom solutions. One of the most popular options is to employ a birdcage MR coil which forms a cylinder of rungs that loop around the head (Bestmann et al., 2003; Dowdle et al., 2018; Hanlon et al., 2016). The birdcage MR coil is large enough to allow the TMS coil to be placed on the subject's head but it provides relatively poor signal-to-noise ratio (SNR) due to the large distance between the coil and the head.

Other studies have used a dedicated 7-channel MR coil array that directly attaches onto the TMS coil and sits between the head and the TMS coil in a thin case (Navarro de Lara et al., 2015, 2017). This solution offers very high SNR at the location of stimulation but low SNR far from the stimulation location (it also requires the purchase of the 7-channel MR coil, which is quite expensive). It is possible to use a second 7-channel coil on the other side of the head to increase the SNR but the purchase of two additional MR coils becomes prohibitively expensive for many imaging centers. Additionally, the use of the 7-channel coil substantially weakens the strength of the TMS pulse that reaches the brain because the MR coil introduces a sizeable gap between the TMS coil and the head. This gap necessitates the use of much higher TMS intensities, which, in turn, can cause the TMS coil to overheat more quickly.

Here, we develop a fully adjustable, MRI-safe TMS stand that can support a range of MR coils, TMS coils, and neuronavigation systems. It also features a fully adjustable mirror attachment. The stand drastically lowers the difficulty associated with positioning and stabilizing the TMS coil, and allows precise targeting with the help of neuronavigation. The stand uses ball-and-socket joints that can tighten into place to adjust the positioning of all of the TMS- and MRI-related components. The stand can be built entirely using 3D-printed parts and PVC piping, and is therefore inexpensive. We provide design files for 3D-printed parts as well as instructions for assembling the stand.

Timing TMS with respect to MRI

Another challenge associate with concurrent TMS-fMRI is that it is difficult to time the TMS pulse with respect to the fMRI pulse sequence to avoid image artifacts. Timing the TMS pulse directly from a standard commercially available computer that also controls the stimulus presentation is prone to high variability in the actual timing of the stimulation relative to the fMRI pulse sequence (Caparelli et al., 2020).

To avoid TMS-induced fMRI artifacts, previous studies typically left a delay at the end of each volume with no slice acquisition to send the TMS pulse. However, such delays increase the length of the repetition time (TR), thus allowing less MRI data to be collected and consequently decreasing the ability to find reliable effects in the fMRI data. For example, in one study, a 1-second TR consisted of 680 ms of fMRI acquisition followed by a 320-ms gap, which means that the TR was 47% longer than it would have been without the gap (Navarro de Lara et al., 2017). To maximize the signal acquired, other studies did not introduce a gap but instead delivered the TMS pulses during echo-planar imaging (EPI) slices of low interest (Rafiei et al., 2021). Delivering TMS during specific EPI slices allows researchers to have the shortest possible TR with the only signal loss coming from the targeted slices (which need to be removed due to the TMS-induced artifacts). Unfortunately, variable delay in the TMS stimulation can occur when a commercial computer is used to control timing of the TMS pulse thus leading to artifacts that corrupt slices other than the intended ones. There are not readily available methods for implementing a TMS-fMRI interface with precise TMS timing.

To address this issue, we develop an interface that precisely times the TMS pulse relative to the fMRI pulse sequence using a microcontroller. The circuitry for the device is laid out on a printed circuit board (PCB). The device connects to a computer (which can be used to control the stimulus presentation) via USB, to the MRI scanner via an optical cable, and to the TMS machine via a 9-pin COM port. The interface produces a delay with a variance of under 1 ms which is likely primarily due to variation in the actual TMS device. We provide PCB design files, software, and instructions for producing this tool.

Tracking TMS coil placement in the MRI

A third challenged of concurrent TMS-fMRI is that it is difficult to establish the precise placement of the TMS coil on the subject's brain inside the scanner, as well as to track if the TMS coil is displaced over the duration of the experiment due to subject movement. Because of this difficulty, most studies do not localize the TMS inside the scanner even though there is significant variability when placing the TMS coil. Several publications use in-scanner marks to localize the TMS coil location from the T1-weighted image (de Weijer et al., 2014; Kemna &

Gembris, 2003; Rafiei et al., 2021) but instructions for setting up an in-scanner localization tool are not available.

To address this issue, we developed a procedure that uses Vitamin E capsules, placed equidistant from the center of the TMS coil, to localize TMS coil placement inside the MRI scanner. We provide code that can be used to find the exact position of the TMS from the T1-weighted image using the signal generated by the Vitamin E capsules.

**Setup and operation**

3D-printed stand

We designed a 3D-printed stand that allows the TMS coil, MR coils, and mirror to be seamlessly placed together inside the scanner (**Figure 1**). The TMS stand contains no metal and is MRI safe. The stand can support a range of MR coils, TMS coils, and neuronavigation systems. However, the stand was designed for and tested on the MagVenture MRi-B91 TMS coil (https://www.magventure.com/us/tms-research/products-overview/research-coils/coils/mri-b91), Localite Neuronavigation system (https://www.localite.de/en/products/tms-navigator/), Siemens 4-channel large flex coil (https://www.siemens-healthineers.com/en-us/magnetic-resonance-imaging/options-and-upgrades/coils/4-channel-flex-coils) and the bottom half of the Siemens 20-channel head coil (https://www.siemens-healthineers.com/en-us/magnetic-resonance-imaging/options-and-upgrades/coils/32-channel-head-coil). Constructing the stand is done in three simple steps: (1) print all of the 3D parts using the provided files in the GitHub repository (https://github.com/samuelgoldsteinGT/concurr_TMS_fMRI_Tools), (2) acquire PVC piping of the correct thickness and cut it to the correct length using information on PVC specifications found in the GitHub repository, and (3) glue the 3D-printed parts and PVC pipes together.

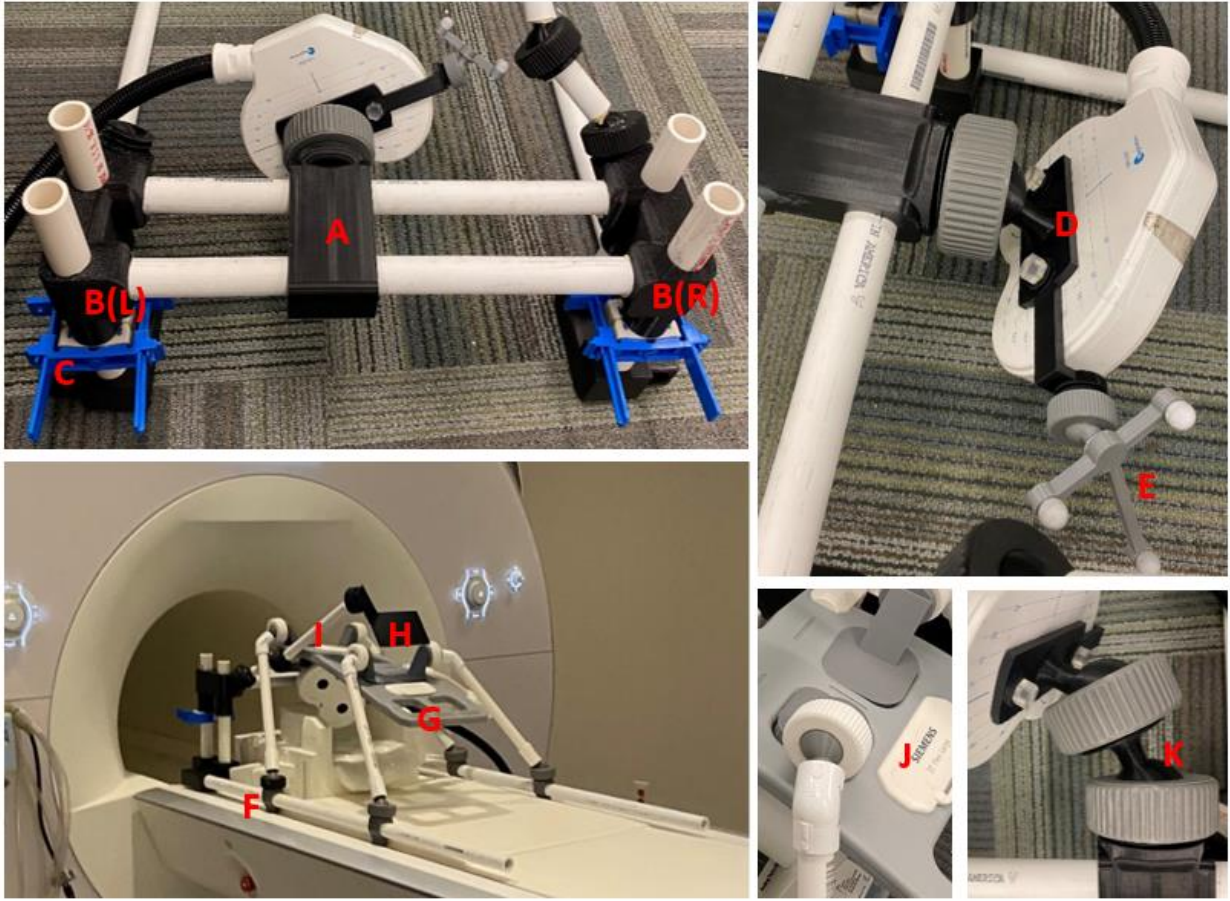

**Figure 1. TMS stand**. The stand integrates the TMS coil, MR coils, and mirror into a unified system. The location of the base of the TMS arm can be adjusted in all three dimensions with the following features: Horizontal slider (A) moves along a horizontal track. Vertical slider (B) moves along a vertical track and can be secured using a clamp (C). A ball-and-socket joint screws on to the TMS coil (D) and can be used to adjust the approach angle of the TMS. The attachment (E) can be used by the neuronavigation system to track the positioning of the TMS. Its angle is adjustable via a ball-and-socket joint. The stand can attach to the track on the MRI bed (F). The MR flex coil (G) can be adjusted via four arms will ball-and-socket joints that attach to the stand and can slide along the track (F). A mirror holder (H) can be positioned using a fully adjustable arm (I) with ball-and-socket joints that tightens into place. The arms of the flex coil holder have attachments (J) that snap into place on the flex coil. An arm extension (K) can be placed on the TMS arm so that a wider range on angle can be reached.

3D-printed parts were designed in Autodesk Fusion 360, prepared for printing using Ultimaker Cura, and printed using Ultimaker S3 3D printer. The 3D design files (.stl files) can be opened on Ultimaker Cura and printed on a wide range of 3D printers. As some functional parts require more strength than others, we provide a 3D design specification file with the recommended infill density and wall thickness. Additionally, we provide quantities for each 3D printed parts.

PVC piping can be purchased at a home improvement store (e.g., Home Depot or Lowes in the USA). To get the correct dimensions for the PVC piping, one can purchase piping of the correct thickness and use a tubing cutter tool, which can be found on Amazon or a home improvement store, to cut the piping to the correct length. PVC piping used for the stand is either has a 0.84-inch outer diameter or a 1.05-inch outer diameter. We provide a file containing required PVC tubing length and thickness for the frame of the stand. Also, we provide quantities of all PVC parts.

All the parts can be attached together with super glue (see **Figure 1** for pictures of the fully constructed stand). Stunt clamps, the blue part in **Figure 1C**, clamp on to the vertical PVC piping and ensure that the stand does not slide down from the weight of the TMS coil and cable. They can be purchased at a home improvement store. For the neuronavigation attachment for the TMS stand (**Figure 1E**), reflective reference balls need to be superglued onto the neuronavigation tripod 3D printed part (spare reference balls are typically provided by neuronavigation manufacturers such as Localite).

The stand attaches to the MRI bed via the base of the flex coil arm (**Figure 1F**). For more secure attachment to the bed, the PVC tubing that runs along length of bed can be clipped on to the cord holder accessory that run along the track.

The stand features a mirror holder that can easily be adjusted via ball-and-socket joints to allow the subject to see a monitor displaying the cognitive task. There are two variations of the mirror holder (**Figure 1H**), one that links up to the right vertical track base (**Figure 1B(R)**), the 3D printed part that allows the TMS coil holder to slide up and down the vertical PVC track and one that hooks up to the left vertical track base (**Figure 1B(L)**). If the TMS coil is targeting in area in the right hemisphere, it is easier to use the mirror attachment on the left side so that the coil does not interrupt that placement of the mirror arm. An MRI-safe mirror can be attached to the mirror holder via Velcro.

The 3D-printed stand allows the TMS coil and MR coils to be placed in a large number of configurations to achieve the desired outcome for concurrent TMS-fMRI studies with few additional resources. The stand allows several aspects of the concurrent TMS-fMRI setup to be efficient.

First, the stand allows the TMS coil to quickly be positioned in the correct location due to adjustability in all three spatial directions and all approach angles due to ball-and-socket joints. The ball-and-socket joint tightens into place and the TMS coil placement remains stable during experimentation. The ball-and-socket joints are also fully detachable for easier portability. The neuronavigation attachment with the three reflective balls is adjustable via a ball-and-socket joint to allow the neuronavigation to track the TMS location at different locations and angles (**Figure 1E**).

Second, the stand allows the Siemens 4-channel large flex coil to quickly be adjusted. The flex coil is made with a flexible material that allows it to be adjusted close to the subject's head without touching the TMS coil. The optimal position of the flex coil depends on the TMS coil placement. For example, the flex coil can be placed closer to the head if the TMS flex coil is targeting a region in the parietal cortex rather than the frontal cortex. Placing the flex coil closer to the head allows for better SNR.

Third, the mirror can be quickly adjusted to allow the subject to see the screen though the holes in the flex coil. The mirror attaches to an adjustable arm with ball-and-socket joints to allow for optimal placement.

Timing interface

The timing interface facilitates communication among the TMS, MRI, and PC with task-presentation software and precisely times the TMS pulse to be interleaved with the fMRI pulse sequence at a predefined slice and delay. Building and setting up this tool can be completed in three major steps. First, one must send the printed circuit board (PCB) design files to a manufacturer to print the board and order electronic parts from an electronics store. Second, one must fuse electronic parts to the board via soldering, a process where metal contacts from the electronic parts are melted to the electrical contact of the board with a metal alloy and a hot iron. Third, one can then simply download the provided code for operating the interface to the board.

We provide PCB design files, called Gerber files and Drill files, in the GitHub repository which can be sent directly to manufacturers. The manufacturer must receive all the Gerber files and Drill files. We provide the timing interface PCB (**Figure 2**), the circuit schematic (**Supplementary Figure 1**) and PCB layout (**Supplementary Figure 2**). For the PCB shown (**Figure 2**), we sent the design files to a PCB manufacturing company named JLCPCB (jlcpcb.com) that has a production time of about 3 days. We provide a parts list of the items that are to be soldered on the PCB on the GitHub repository. We ordered all parts from Digi-Key electronics (digikey.com) and Amazon (amazon.com).

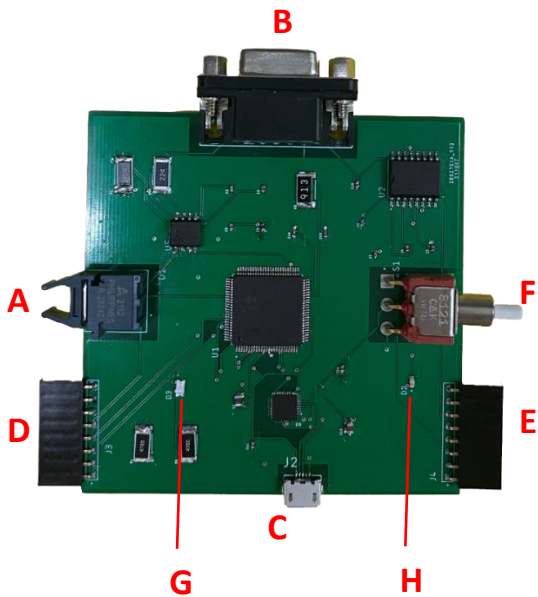

**Figure 2. Timing interface**. This device times the TMS pulse based on MRI optical trigger from the slice timing and the PC running the task-presentation software (A) Connection for optical trigger cable which connects to MRI optical trigger cable. (B) 9-pin serial cable connection that connects to COM 2 port on TMS. (C) Micro USB connection for powering board and communicating with PC. (D) Header connectors for connecting to the MSP-EXP432P401R evaluation board to program interface. (E) Header connector for accessing nodes of the circuit. (F) Reset button. (G) Green LED. (H) Red LED.

Electronic parts are connected to the board via surface mount soldering, a type of soldering where parts are mounted to the surface of the PCB. Most electrical engineering departments of universities have the resources and expertise to mount all the electronic components to the board.

For the interface in this paper, the board was soldered at Georgia Tech Hive Makerspace, a lab at Georgia Tech that gives student access to tools such as soldering irons and supplies such as capacitors. Materials required are a soldering iron, solder wire, and flux which is a chemical cleaning agent for solder.

The microcontroller (that is, the computer processor that connects to all the circuits in the timing interface) is a Texas Instruments MSP 432. The code that runs on the microcontroller is written in C and is provided in the GitHub repository. The C code must be placed into a project in Code Composer Studio (https://www.ti.com/tool/CCSTUDIO), an integrated development environment software that runs on a personal computer and is used to program microcontrollers such as the MSP 432. MSP432 driver library, a C library containing relevant functions for programming an MSP 432 microcontroller, must be implemented with Code Composer Studio for the code to compile.

An MSP-EXP432P401R evaluation board (which can be purchased at any electronics supplier such as Digi-Key) is used to connect timing interface to a PC running code Composer Studio. The MSP-EXP432P401R evaluation board connects to the PC via USB. The evaluation board connects to the timing interface via the header pins on the Timing Interface (**Figure 2D**). Once the systems are connected, the C code should be uploaded to the timing interface with the debugger command. If the red LED on the timing interface turns on when the push button is pressed (**Figure 2F**), the timing interface is successfully programmed. After this action, the timing interface can operate without Code Composer Studio as long it is connected to a power supply via micro-USB (**Figure 2C**).

When using the provided C codes, the MRI must be set to trigger via optical cable at every slice rather than every TR. Alternatively, the C codes can be modified so that they work when the MRI trigger is received once per TR. More detailed instructions for operating the interface can be found in the Supplementary Material.

Coil localization tool

A critical issue for any concurrent TMS-fMRI experiment is how to accurately localize the exact location on the brain where TMS was actually delivered. Accurate localization of the TMS coil enables a researcher to pinpoint the place of maximum stimulation, and also allows estimating the coil displacement over the course of an experiment by comparing the anatomical images collected at the beginning and end of a scanning session. There is currently no standard tool for in-scanner localization of the actual TMS coil position. Here we provide instructions for placing Vitamin E capsules on the TMS coil and code that uses the signal from these Vitamin E capsules to localize the TMS coil using information from a T1-weighted structural image. A similar procedure was implemented in our recent concurrent TMS-fMRI study (Rafiei et al., 2021).

The Vitamin E capsules need to be attached to the TMS coil in a way that allows the determination of the exact TMS coil location with respect to a subject's brain. The arrangement that we recommend consists of six Vitamin E capsules on the surface of TMS coil and one capsule to the surface behind it (**Figure 3**). In this arrangement, the capsules are taped to the surface of the coil that contacts the head and are arranged in a hexagon shape such that the center of the hexagon is in the middle of TMS coil. Further, an extra Vitamin E capsule is taped at the center of TMS coil but on the surface behind it. This arrangement makes it easy to find the line perpendicular to the TMS coil where the induced magnetic field is known to be maximal (de Weijer et al., 2014). We call the point on the head where the center of the TMS coil touches the "entry point" and the line perpendicular to the TMS coil the "entry line."

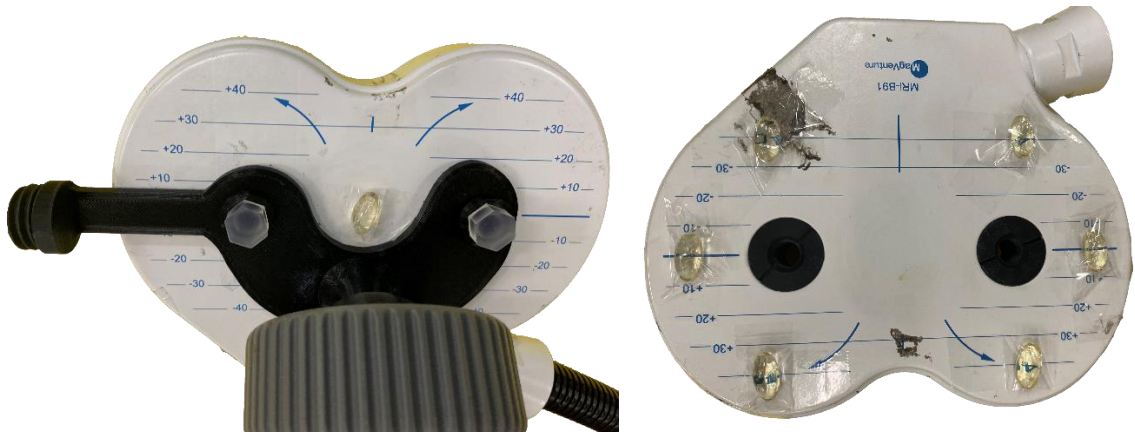

**Figure 3. Vitamin E placement for TMS coil localization**. Six Vitamin E capsules are arranged on the plane that contacts the head equidistant from the center. Another Vitamin E is placed in the center of the TMS coil on the other side.

We provide code to compute the entry point and entry line using the location of the Vitamin E capsules on a T1-weighted structural image. The code receives the location of the capsules in structural image space as input. This step needs to be done manually by the user who has to identify all seven Vitamin E capsule locations on the T1-weighted image and select the location that corresponds to the capsule that is attached behind the coil. The code then first computes the equation of the TMS coil surface that touches the subject's head (we call this the "stimulation surface") (**Figure 4**). It then determines the entry line as the line which is perpendicular to the stimulation surface and passes through the Vitamin E capsule attached at the midpoint behind the TMS coil. The entry point is then computed by either finding the intersection of the entry line and the stimulation surface or by finding the average point of all Vitamin E locations attached on the stimulation surface. Finally, the code allows a user to define the stimulation spot as the point on entry line which makes first contact with the brain tissue by providing a visualization tool based on the MarsBaR toolbox in SPM (Brett et al., 2002).

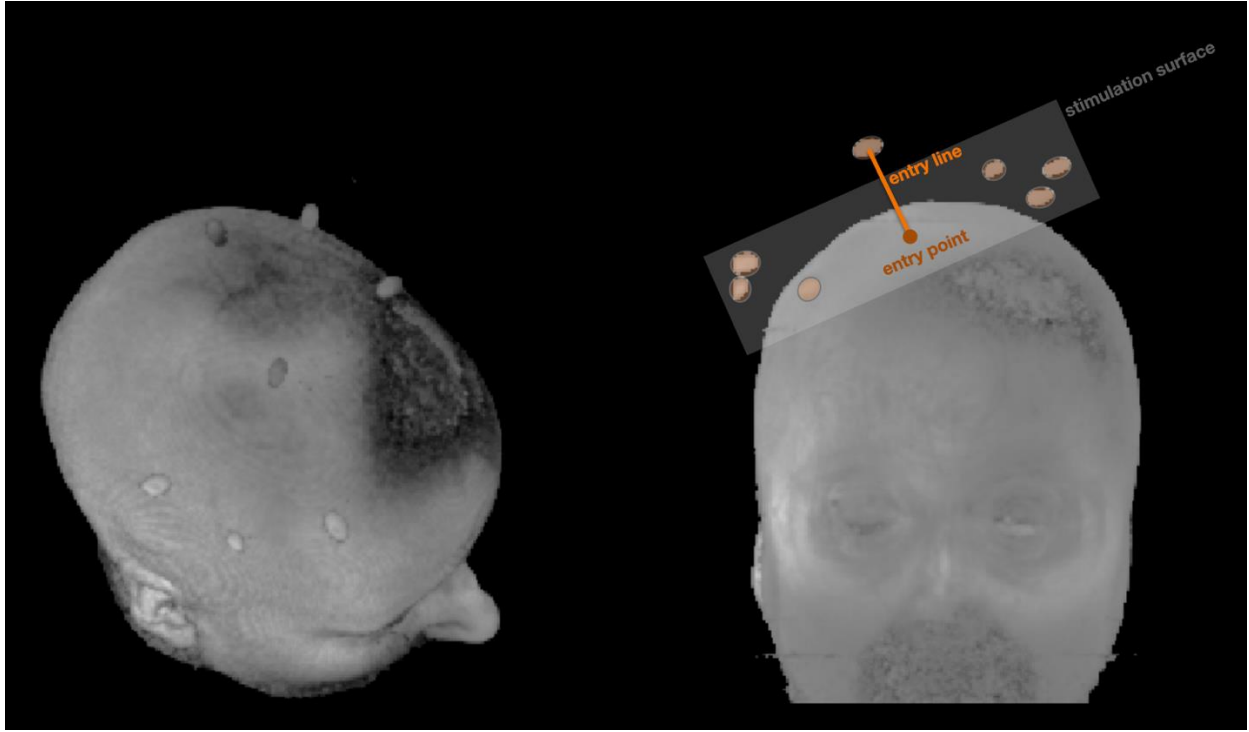

**Figure 4. Visualization of Vitamin E capsules inside MRI scanner for an example subject**. Hexagonal arrangement of Vitamin E capsules on the surface of the coil is evident on the structural image. The stimulation surface is the surface which passes through the six Vitamin E capsules taped to the side of the TMS coil that makes contact with the subject's head. The entry line passes through the remaining Vitamin E capsule (taped on the opposite side of the TMS coil) and is perpendicular to the stimulation surface. The entry point is defined as the place of intersection of the stimulation surface and the entry line. Finally, the stimulation spot could be determined by the user by locating the point on the entry line that first makes contact with the brain tissue.

## Validation

<u>3D-printed stand</u>

To test the adjustability of the stand, we used it to place the TMS coil over several key brain areas: frontal eye fields, ventromedial prefrontal cortex, dorsolateral prefrontal cortex, and primary motor cortex using Localite Neuronavigation (**Figure 5**). Each region could be localized in under 5 minutes.

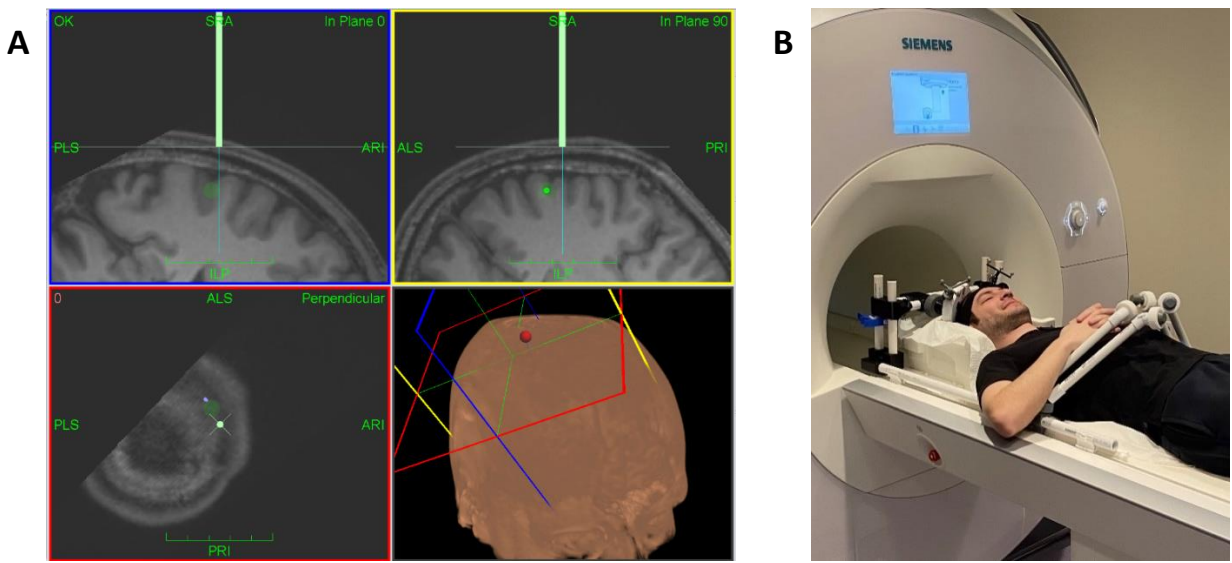

**Figure 5. TMS Neuronavigation.** Localizing the right frontal eye fields (FEF) via Localite neuronavigation.

We further conducted a resting state scan on a healthy, right-handed male with the full TMS stand setup including the presence of the TMS coil (but without sending TMS pulses). Data collection was performed on a Siemens Magnetom Prisma 3T scanner. We collected a T2*-weighted gradient echo-planar image sequence with partial brain coverage near the top of the head (TR = 1000 ms, 25 descending slices, FoV = 220 mm; TE = 40 ms; flip angle = 50°; voxel size = 3.44 x 3.44 x 5 mm$^3$). In addition, we collected a high-resolution T1-weighted image (FoV

= 256 mm; TE$_{1-4}$ = 1.69 - 7.27ms; TR = 2530 ms; 176 slices; flip angle = 7°; voxel size = 1.0 x 1.0 x 1.0 mm$^3$). We first examined the T1-weighted image, which did not reveal any artifacts related to the presence of the TMS stand. We then examined the temporal signal-to-noise ratio (tSNR) obtained in different parts of the brain using our setup. Not surprisingly, there was higher tSNR in the occipital lobe because of its proximity to the bottom of the 20-channel coil (**Figure 6**). There was relatively lower but still excellent tSNR in the frontal lobe that is closest to the 4-channel flex coil positioned near the forehead of the subject. Indeed, a previous review paper pertaining to tSNR values for fMRI studies reported that typical tSNR values are in the range of 10 to 50 (Welvaert & Rosseel, 2013), which matches the range observed in the frontal lobe using our setup.

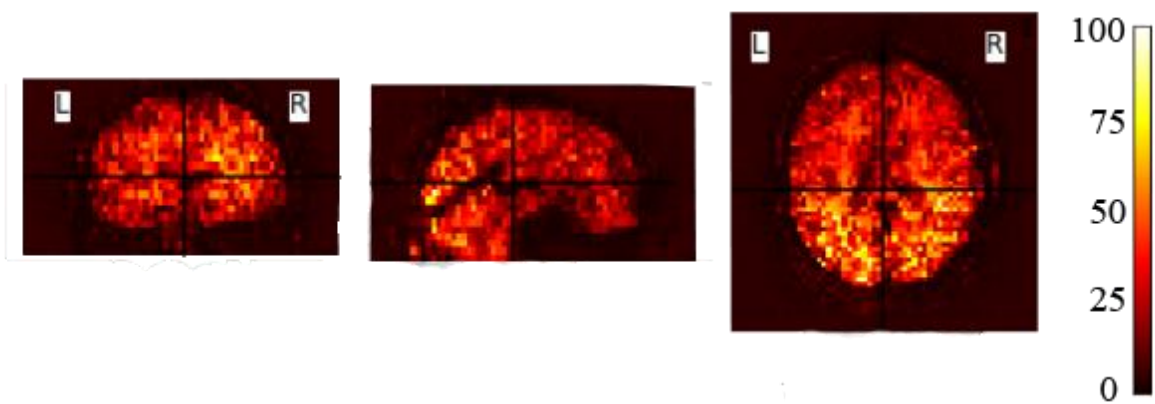

**Figure 6. tSNR brain map for a resting state scan.** The figure plots tSNR values for each voxel for the same subject from Figure 5. Excellent tSNR is obtained in the whole brain.

Timing interface

We conducted a simulation with 50 trials to determine the delay between the interface receiving a trigger from the MRI optical cable and the TMS producing a pulse. A debouncing switch was used to simulate the MRI trigger from a slice. A hall effect magnetic field sensor captured the magnetic field generated from the TMS. An oscilloscope captured the timing of the magnetic field and simulated the MRI trigger (**Supplementary Figure 3**). We found a mean delay of 4.63 ms with variance of 0.84 ms (**Figure 7**). The small variance in measured delay times means that it is easy to ensure that TMS is delivered with very high precision that substantially exceeds the precision achieved when using commercial computers for stimulation.

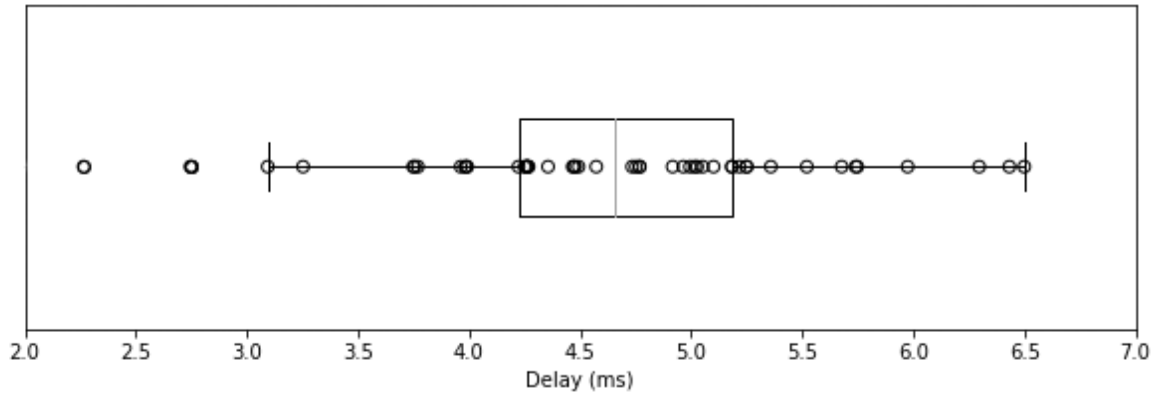

**Figure 7. Delays for timing interface.** The distribution of the delays between the interface sending instructions to the TMS and the TMS evoking a magnetic field.

## Discussion

Concurrent TMS-fMRI experiments are uniquely able to reveal whole-brain dynamics based on a causal, experimenter-controlled input. However, there is a large logistical barrier involved in concurrent TMS-fMRI experimentation. In this paper, we introduce three tools to address this barrier.

First, we introduce a 3D-printed TMS stand that makes the experimental setup much more efficient. The TMS coil can localize all brain areas that are not obstructed by the placement of the MR coils such as the occipital lobe. Compatibility with the Siemens flex coil allows for high signal-to-noise ratio to be achieved by maximizing its proximity to the head without touching the TMS coil. Second, we introduce a timing interface that allows the experimenter to have much better control on when the TMS protocol is executed relative to the fMRI pulse sequence. Third, we introduce a system for localizing the TMS coil positioning during the T1 scan via the placement of Vitamin E capsules on the TMS coil.

A limitation of the 3D-printed stand is that it is designed to be compatible specifically for MagVenture MRi-B91 TMS coil, Localite Neuronavigation system, and Siemens 4-channel large flex coil. Future directions for the stand would be redesigning the 3D-printed parts to accommodate a wider range of TMS and MR coils, as well as neuronavigation systems. Researchers who use other tools can contact the first author for assistance with designing custom stands that tailored towards the available equipment.

A future direction for the 3D-printed stand is designing an automatic positioning system using cables, pullies, and real-time feedback from the neuronavigation system to automatically position the TMS coil on the desired region of interest. There are currently no tools for automatically positioning TMS inside the MRI. The system controlling the movement of the stand can be made MRI-safe and could be controlled by motors that are placed a safe distance away from the MRI. In addition to quickly positioning the TMS coil on the subject's head during setup, this tool would allow the TMS target to change inside the MRI during an experiment without taking the subject out of the scanner.

Kemna, L. J., & Gembris, D. (2003). Repetitive transcranial magnetic stimulation induces different responses in different cortical areas: A functional magnetic resonance study in humans. *Neuroscience Letters*, *336*(2), 85–88. https://doi.org/10.1016/S0304-3940(02)01195-3

Loo, C. K., & Mitchell, P. B. (2005). A review of the efficacy of transcranial magnetic stimulation (TMS) treatment for depression, and current and future strategies to optimize efficacy. *Journal of Affective Disorders*, *88*(3), 255–267. https://doi.org/10.1016/j.jad.2005.08.001

Navarro de Lara, L. I., Tik, M., Woletz, M., Frass-Kriegl, R., Moser, E., Laistler, E., & Windischberger, C. (2017). High-sensitivity TMS/fMRI of the Human Motor Cortex Using a Dedicated Multichannel MR Coil. *NeuroImage*, *150*, 262–269. https://doi.org/10.1016/j.neuroimage.2017.02.062

Navarro de Lara, L. I., Windischberger, C., Kuehne, A., Woletz, M., Sieg, J., Bestmann, S., Weiskopf, N., Strasser, B., Moser, E., & Laistler, E. (2015). A novel coil array for combined TMS/fMRI experiments at 3 T. *Magnetic Resonance in Medicine*, *74*(5), 1492–1501. https://doi.org/10.1002/mrm.25535

Paus, T., & Barrett, J. (2004). Transcranial magnetic stimulation (TMS) of the human frontal cortex: Implications for repetitive TMS treatment of depression. *Journal of Psychiatry and Neuroscience*, *29*(4), 268–279.

Rafiei, F., Safrin, M., Wokke, M. E., Lau, H., & Rahnev, D. (2021). Transcranial magnetic stimulation alters multivoxel patterns in the absence of overall activity changes. *Human Brain Mapping*, *42*(12), 3804–3820. https://doi.org/10.1002/hbm.25466

# Supplementary Materials

**Timing interface operation**

After connecting the interface to the TMS, MRI, and PC, the reset button should be pressed. This will cause the red LED light up. When you start the fMRI sequence, the red LED should turn off.

Before starting the fMRI sequence, there are three serial commands to send to the interface to configure it. Set the number of slices per TR via [254, 111, number of slices, 255]. Set which slice to stimulate starting from 1 to number of slices via [254, 112, slice to stimulate, 255]. Set how many milliseconds to wait before stimulating via [254, 113, delay in milliseconds, 255]. Note that this delay is in addition to the delay that the TMS and interface already produces.

The next set of commands are specific to the MagVenture TMS. The identical serial commands that would be sent directly to the COM2 port on the TMS are to be sent to the interface: enable TMS via [254, 3, 2, 1, 0, 139, 255], set amplitude via [254, 3, 1, Amplitude, 0, CRC-8 of amplitude, 255] where code for 8-bit cyclic redundancy check (CRC-8) is in the GitHub corresponding to this paper, send a single TMS pulse via [254, 3, 3, 1, 0, 32, 255] and a pulse train via [254, 3, 4, 0, 0, 158, 255]. The pulse train is configured directly on the TMS. The green LED of the interface will briefly turn on when a pulse command is received as it waits for the fMRI trigger. All of the serial signals to the interface must be sent at a 38400 baud rate with no parity and one stop bit. Example Matlab code for communicating with the interface is provided in the GitHub repository.

During operation, the interface sends 0x69 serial bit stream via a serial communication protocol called Universal Asynchronous Receiver/Transmitter (UART) back to the PC for every fMRI slice. Also, it sends 0x74 serial bit stream via UART back to the PC whenever a TMS pulse is executed. These serial codes can be used to time the stimulus presentation controlled by the PC, relative to the operation of the TMS. This communication is crucial when timing the TMS protocol during a concurrent task.

A

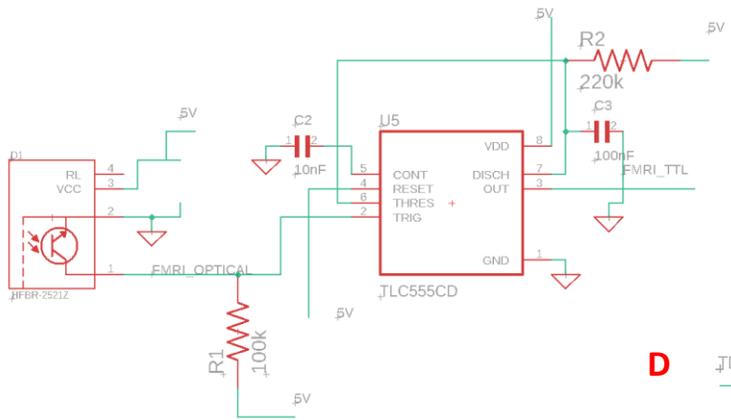

B

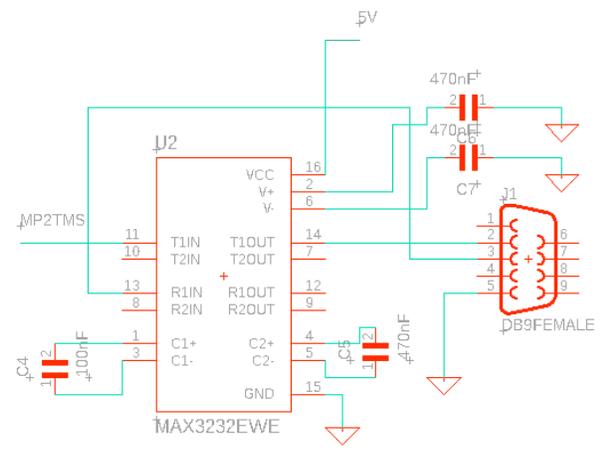

D

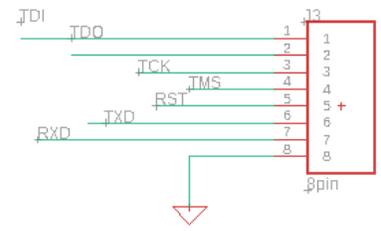

E

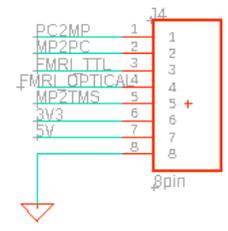

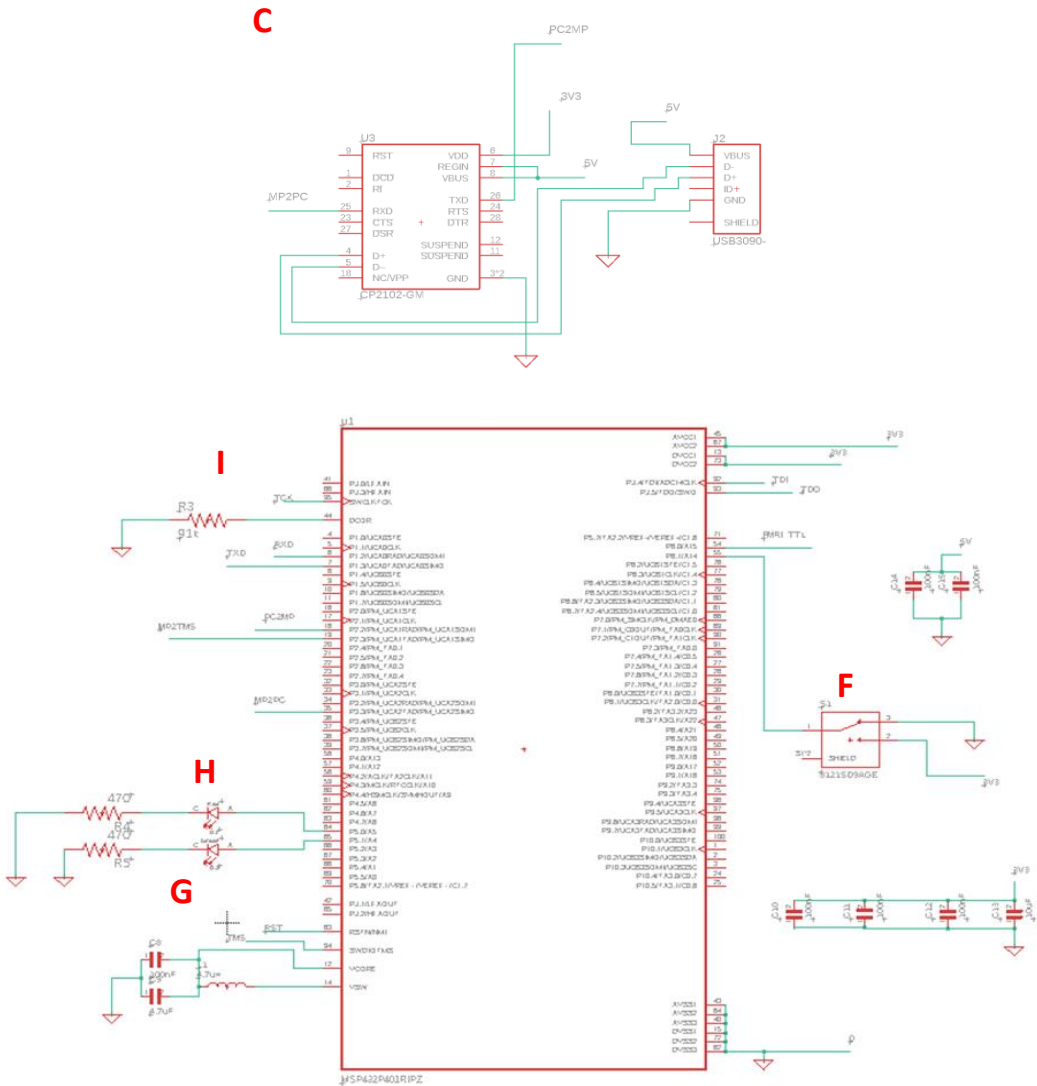

**Supplementary Figure 1. Circuit schematic for timing interface.** (A) Circuit for processing optical cable input from fMRI. (B) Circuit for sending signal to TMS via the 9-port serial port. (C) Circuit for interfacing with micro-USB. (D) Header connector pinout for programming. (E) Header connector pinout for accessing different nodes of circuit. (F) Circuit for reset button. (G) Green LED. (H) Red LED. (I) MSP432 microprocessor with connections.

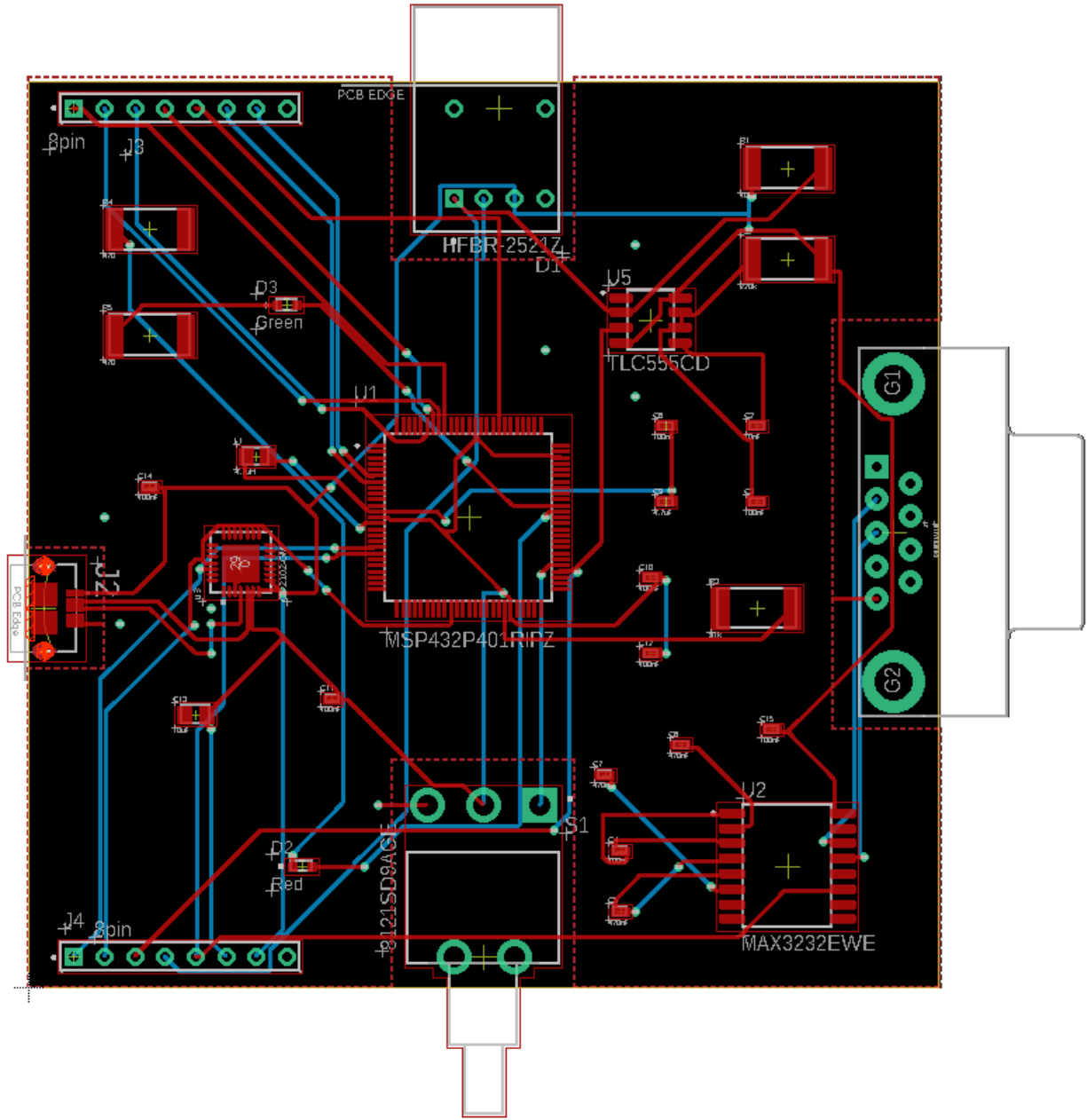

**Supplementary Figure 2. PCB layout.** Layout for the printed circuit board that has to be sent to the manufacturer for printing. The layout contains all the circuitry for the timing interface. After receiving the board, the electronic parts must be soldered on it.

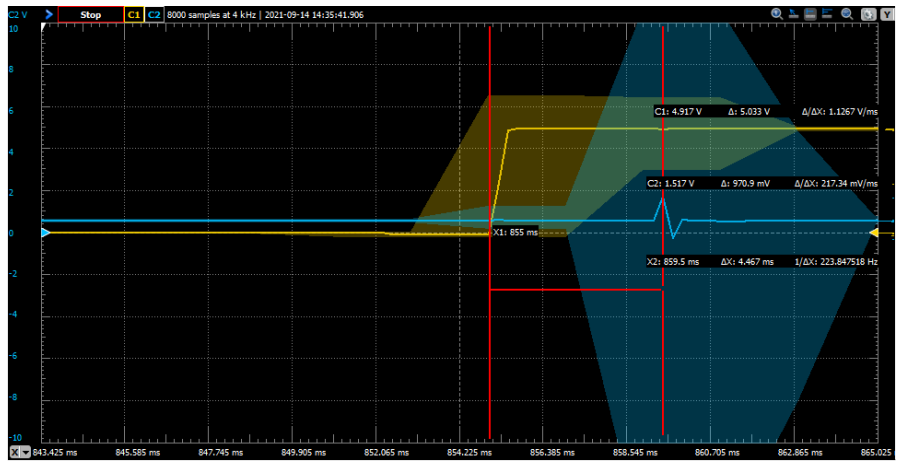

**Supplementary Figure 3. Simulation for timing interface.** Oscilloscope readings for fMRI slice simulation for one of the trials. Channel 1 (in yellow) simulated fMRI slice trigger signal using a debouncing switch. Channel 2 (in blue) depicts the voltage readout from a hall effect magnetic field sensor placed near the TMS coil. The delay between the trigger input and TMS pulse is 4.47 ms in this trial.